\documentclass[12pt,a4paper,dvips]{article}
\usepackage{a4p}
\usepackage{cite,mcite}
\usepackage{graphicx}
\usepackage{physics }
\usepackage{l3_title,ifthen}

\journalname{Phys. Lett. B}

\date{June 20, 2003}

\preprint{2003-034}

\newlength{\capindent}
\newlength{\capwidth}
\newlength{\figwidth}
\newcommand{\icaption}[2][!*!,!]{\hspace*{\capindent}%
  \begin{minipage}{\capwidth}
    \ifthenelse{\equal{#1}{!*!,!}}%
      {\caption{#2}}%
      {\caption[#1]{#2}}
  \end{minipage}}
\newcommand{\pho}{\phantom{0}}
\newcommand{\SM}{Standard Model}

\newcommand{\MC}{Monte Carlo}

\newcommand{\ZZ}{\ensuremath{\mathrm{ZZ}}}

\newcommand{\qqll}{\ensuremath{\rm q \bar q\ell^+\ell^- }}
\newcommand{\llnn}{\ensuremath{\rm \ell^+\ell^- \nu\bar\nu}}

\newcommand{\llll}{\ensuremath{\rm \ell^+\ell^-\ell^{\prime +}\ell^{\prime -} }}
\newcommand{\qqnn}{\ensuremath{\rm q\bar q\nu\bar\nu}}
\newcommand{\qqqq}{\ensuremath{\rm q\bar qq^\prime\bar{q}^\prime}}

\newcommand{\qqee}{\ensuremath{\rm q\bar q e^+ e^-}}
\newcommand{\qqmm}{\ensuremath{\rm q\bar q \mu^+ \mu^-}}
\newcommand{\qqtt}{\ensuremath{\rm q\bar q \tau^+ \tau^-}}

\newcommand{\bbll}{\ensuremath{\rm b\bar b \ell^+ \ell^-}}

\newcommand{\ZZtobbX}{\ensuremath{\rm ZZ \rightarrow \mathrm{ b \bar b X}}}
\newcommand{\eeto}{\ensuremath{e^+ e^- \rightarrow \;}}
\newcommand{\bbnn}{\ensuremath{\mathrm{b\bar b \nu \bar\nu}}}
\newcommand{\qqbb}{\ensuremath{\mathrm{q\bar q b \bar b}}}

\begin{document}
\begin{titlepage}
\title{Z Boson Pair-Production at LEP}
\author{The L3 Collaboration}
%
% The abstract
%
\begin{abstract}

Events stemming from the pair-production of Z bosons in $\rm e^+e^-$
collisions are studied using 217.4\,\pb\ of data collected with the L3
detector at centre-of-mass energies from $200\GeV$ up to
$209\GeV$. The special case of events with b quarks is also investigated.

Combining these events with those collected at lower centre-of-mass
energies, the Standard Model predictions for the production mechanism are verified.
In addition, limits are set on anomalous couplings of
neutral gauge bosons and on effects of extra space dimensions.

\end{abstract}

\submitted
\end{titlepage}

%
%%%%%%%%%%%%%%%%%%%%%%%%%%%%%%%%%%%%%%%%%%%%%%%%%%%%%%%%%%%%%%%%%%%%%%%%%%%%%%%
\section{Introduction}
%%%%%%%%%%%%%%%%%%%%%%%%%%%%%%%%%%%%%%%%%%%%%%%%%%%%%%%%%%%%%%%%%%%%%%%%%%%%%%%
%

The pair-production of Z bosons in $\rm e^+e^-$ collisions at LEP was
observed~\cite{l3zz183} with the L3
detector~\cite{l3_00,*l3_55,*l3lumi,*l3alr,*l3smd,*l3fb,*l3egap} once
the accelerator centre-of-mass energy, $\sqrt{s}$, exceeded the
production threshold of $2m_{\rm Z}$, where $m_{\rm Z}$ denotes the Z boson
mass. Numerous studies from data
samples collected at the steadily increasing $\sqrt{s}$ and integrated luminosities were
reported by L3~\cite{l3zz189,l3zz196} and other
collaborations~\cite{aleph,*delphi1,*delphi2,*opal}.

In the Standard Model of the electroweak
interactions~\cite{sm_glashow,*sm_salam,*sm_weimberg,*v1,*t1,*v2,*v3},
the Z pair-production is described at the lowest order by two
$t$-channel Feynman diagrams with an internal electron\footnote{In
this Letter, the word electron is used for both electrons and
positrons.} leg, collectively denoted as NC02. A wider definition is
used in this Letter: all diagrams leading to two fermion-antifermion
pairs are considered and kinematic restrictions which enhance the NC02
contribution are enforced. Results in the NC02 framework are also
given.

The study of Z pair-production offers a further test of the Standard
Model in the neutral gauge boson sector and is of particular relevance
as this process constitutes an irreducible background in the search
for the Standard Model Higgs boson at LEP. Events with Z boson
decaying into b quarks have a signature similar to those originated by
the process $\rm e^+e^- \rightarrow ZH\rightarrow
f\bar{f}b\bar{b}$. Their selection and the measurement of their cross
section validate the experimental procedures used in the search of
the Standard Model Higgs boson.

Z pair-production allows the investigation of the anomalous
triple neutral gauge boson couplings ZZZ and
ZZ$\gamma$~\cite{hagiwara,*alcaraz,*gounaris}, forbidden at
tree level in the Standard Model and tests
new theories like possible effects of extra
space dimensions~\cite{agashe,*l3lsg183,*l3lsg189,*melesanchez}.

This Letter describes measurements at two average values of
$\sqrt{s}$, $204.8\GeV$ and $206.6\GeV$, corresponding to
integrated luminosities of 78.5\,\pb{} and 138.9\,\pb{},
respectively. Hereafter, these data samples are denoted as the
$205\GeV$ and $207\GeV$ energy bins. Combined results from the full Z
pair-production sample collected with the L3 detector
are also given in the comparison with Standard Model expectations and
for constraints on  New Physics.

%
%%%%%%%%%%%%%%%%%%%%%%%%%%%%%%%%%%%%%%%%%%%%%%%%%%%%%%%%%%%%%%%%%%%%%%%%%%%%%%%
\section{\MC{} simulations}
%%%%%%%%%%%%%%%%%%%%%%%%%%%%%%%%%%%%%%%%%%%%%%%%%%%%%%%%%%%%%%%%%%%%%%%%%%%%%%%
%

The EXCALIBUR~\cite{exca} \MC{} program is used to model signal and
background neutral-current four-fermion processes. The Z
pair-production process is defined as the subset of the
four-fermion generated phase space satisfying the following kinematics
cuts~\cite{l3zz183,l3zz189,l3zz196}.  The
invariant mass of fermion-antifermion pairs is required to be between $70\GeV$ and $105\GeV$. For events with two identical
pairs, at least one of the possible pairings has to satisfy this
condition. For the $\rm u\bar d d \bar u $, $\rm c\bar s s \bar
c $ and $\rm\nu_\ell\ell^+\bar{\nu_\ell} \ell^-$ ($\rm
\ell=e,\mu,\tau$) final states, the masses of the pairs that could originate from a
W decay have to be either below $75\GeV$ or above $85\GeV$. The polar
angle $\theta_e$ of generated electrons is required to satisfy $|\cos\theta_{\rm e}| < 0.95$.

The Z pair-production cross section is calculated to be $1.07\,$pb and
$1.08\,$pb for the $205\GeV$ and $207\GeV$ energy bins,
respectively. Following a comparison with the GRC4F~\cite{grace} \MC{}
generator, and taking into account the modelling of initial state
radiation, an uncertainty of 2\% is assigned to these calculations.

The cross section for final states with b quark pairs is significantly
smaller. Combining the two energy bins a  cross-section of 0.30\,pb, also with an
uncertainty of 2\%, is expected for an average centre-of-mass energy $\sqrt{s} = 205.9\,\GeV$.

Four-fermion events generated with EXCALIBUR which do not satisfy the
signal definition are considered as background. Background from
fermion pair-production is described by KK2f~\cite{kk2f}
for the processes $\rm e^+ e^- \rightarrow q \overline q (\gamma)$,
$\rm e^+ e^- \rightarrow \mu^+ \mu^- (\gamma)$ and $\rm e^+ e^-
\rightarrow \tau^+ \tau^- (\gamma)$, and BHWIDE~\cite{bhwide} for $\rm
e^+ e^-\rightarrow e^+ e^- (\gamma)$. Background from charged-current
four-fermion processes is generated with EXCALIBUR for the $\rm
e\nu_\e q\bar q'$ final state and with KORALW~\cite{koralw} for W
pair-production and decay in final states not covered by the
simulations listed above. Hadron and lepton production in two-photon processes 
is
modelled by PHOJET~\cite{phojet} and DIAG36~\cite{diag36},
respectively.

The L3 detector response is simulated using the GEANT
program~\cite{geant}, which takes into account the effects of energy
loss, multiple scattering and showering in the
detector. GHEISHA~\cite{gheisha} is used for the simulation of hadronic
interactions. Time dependent detector inefficiencies, as monitored
during the data taking period, are also reproduced.

%
%%%%%%%%%%%%%%%%%%%%%%%%%%%%%%%%%%%%%%%%%%%%%%%%%%%%%%%%%%%%%%%%%%%%%%%%%%%%%%%
\section{Event selection}
%%%%%%%%%%%%%%%%%%%%%%%%%%%%%%%%%%%%%%%%%%%%%%%%%%%%%%%%%%%%%%%%%%%%%%%%%%%%%%%
%

All visible final states of Z pair-production are investigated. For the
$\qqnn$, $\llnn$ and $\llll$ final states, criteria are used which are
similar to those developed at $\sqrt{s}=189\GeV$~\cite{l3zz189} and
$\sqrt{s}=192-202\GeV$~\cite{l3zz196}. For the \qqqq{} and \qqll{}
final states, improved analyses are devised. All selections rely on the
identification of two fermion pairs with masses close to
$m_{\rm Z}$.

Electrons are identified by requiring a well isolated electromagnetic
cluster in the electromagnetic calorimeter with an associated track in the
tracking chamber. To increase efficiency, the track matching
requirement is relaxed in some selections.

Muons are reconstructed from the coincidence of tracks in the muon
spectrometer and the central tracker which are in time with the beam
crossing. Energy depositions in the calorimeters which are compatible
with a minimum ionising particle (MIP) and have an associated track in
the central tracker are also accepted.

Taus are identified by their decays either into electrons or
muons, or into hadrons detected as narrow and isolated low
multiplicity jets associated with one, two or three tracks.

Quark fragmentation and hadronisation yields a high multiplicity of
calorimetric clusters and charged tracks. These are grouped into jets
by means of the DURHAM algorithm~\cite{durham}. The number of
reconstructed jets depends on the thresholds $y_{mn}$ for which a
$m$-jet event is reconstructed as a $n$-jet one.

The tagging of b quarks~\cite{aaron-thesis} relies on the
reconstruction of the decay vertices of weakly decaying b-hadrons with
the silicon vertex detector and the central tracker. The shape and
particle content of the associated jets are also considered.

The four-momenta of neutrinos are derived from all other
particles measured in the event, making use of the hermeticity of the
detector.

%
%%%%%%%%%%%%%%%%%%%%%%%%%%%%%%%%%%%%%%%%%%%%%%%%%%%%%%%%%%%%%%%%%%%%%%%%%%%%%%%
\subsection{The \boldmath{\qqqq{}} channel}
%%%%%%%%%%%%%%%%%%%%%%%%%%%%%%%%%%%%%%%%%%%%%%%%%%%%%%%%%%%%%%%%%%%%%%%%%%%%%%%
%

The study of the \qqqq\ channel~\cite{serge-thesis} starts by selecting high multiplicity
events with a visible energy, $E_{vis}$, satisfying $0.75 <
E_{vis}/\sqrt{s} < 1.35$. The energy imbalance in the directions
parallel and perpendicular to the beam axis have to be
less than $0.2 E_{vis}$ and $0.25 E_{vis}$, respectively. These
criteria suppress fermion-pair production, two-photon interactions and
four-fermion final states with leptons. To further reduce boson
pair-production with leptons, events with isolated electrons or muons with an
energy larger than $40\GeV$ are rejected. Events with isolated photons of
energies above $25\GeV$ are also rejected.

The remaining events are forced into four jets. A kinematic fit
which imposes four-momentum conservation is performed to improve the di-jet
mass resolution.  Among the three possible jet pairings, the pairing $i$
is chosen which minimizes:
\begin{displaymath}
\chi^2_{\rm ZZ} = \left(\Sigma_i - 2 m_{\rm Z}\right)^2 /
\sigma_{\Sigma_{\rm ZZ}}^2 + \Delta_i^2/\sigma_{\Delta_{\rm ZZ}}^2,
\end{displaymath}
where $\Sigma_i$ and $\Delta_i$ are the di-jet mass sum and differences
and $\sigma_{\Sigma_{\rm ZZ}}^2$
and $\sigma_{\Delta_{\rm ZZ}}^2$ their resolutions, determined from 
Monte Carlo. Only events for
which $\Sigma_i > 165\GeV$ are retained.

\begin{figure}
\centerline{\includegraphics*[width=0.6\textwidth]{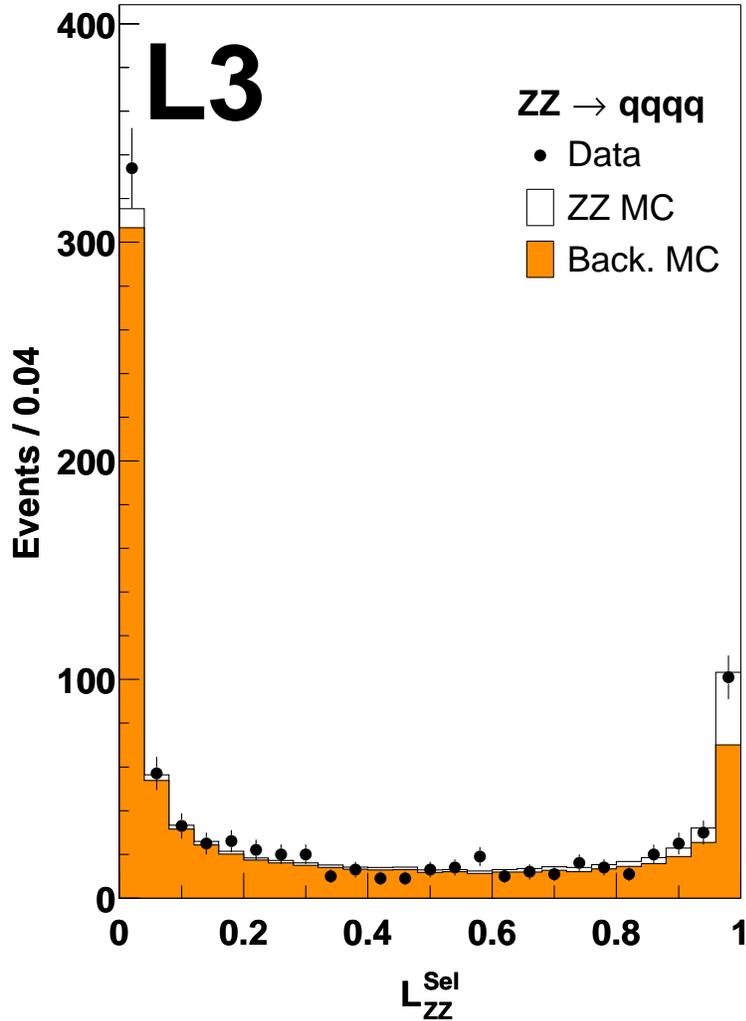}}
\caption{Distribution of the likelihood $L^{Sel}_{\rm ZZ}$ used for
  the \qqqq{} selection. The signal and background Monte Carlo
  distributions are normalised to the expected cross sections.}
\label{fig:qqqq-selection-likelihood}
\end{figure}

The remaining background is formed by events from the $\rm e^+e^-
\rightarrow W^+W^-$ and $\rm e^+e^- \rightarrow q\bar{q}(\gamma)$
processes. A likelihood, $L^{Sel}_{\rm ZZ}$, is built which combines ten
variables: the event sphericity, the largest triple
jet boost~\cite{l3higgs}, the largest jet energy and boost, the largest energy
difference between any two jets, the opening angle between the most
and least energetic jets, $\log y_{34}$, the mass $M_{5C}$ from a
kinematic fit with equal mass constraint, the absolute value of the
cosine of the polar angle of the event thrust
vector,  $| \cos{\theta_T}|$, $\chi^2_{\rm ZZ}$ and the corresponding variable for the
W-pair hypothesis, $\chi^2_{\rm WW}$. The distributions of $L^{Sel}_{\rm
ZZ}$ for data and \MC{} are shown in
Figure~\ref{fig:qqqq-selection-likelihood}. Events with $L^{Sel}_{\rm
ZZ}< 0.1$ are mostly due to the  $\rm e^+e^- \rightarrow q\bar{q}(\gamma)$
process and are not considered in the following.

A second likelihood, $L_{\rm ZZ}$, is built to further exploit the
difference between Z pair-production and the residual background from
W pair-production.  It uses seven variables: $L^{Sel}_{\rm ZZ}$,
$\Sigma_{\rm ZZ}$, the corresponding variable for the W-pair hypothesis,
$\Sigma_{\rm WW}$, the three to four-jet threshold for the JADE clustering
algorithm~\cite{JADE}, the event thrust and the value of the b-tag
variable for the two jets with the highest probability to originate from b
quarks.

\begin{figure}
\begin{center}
\begin{tabular}{cc}
\includegraphics*[width=0.4\textwidth]{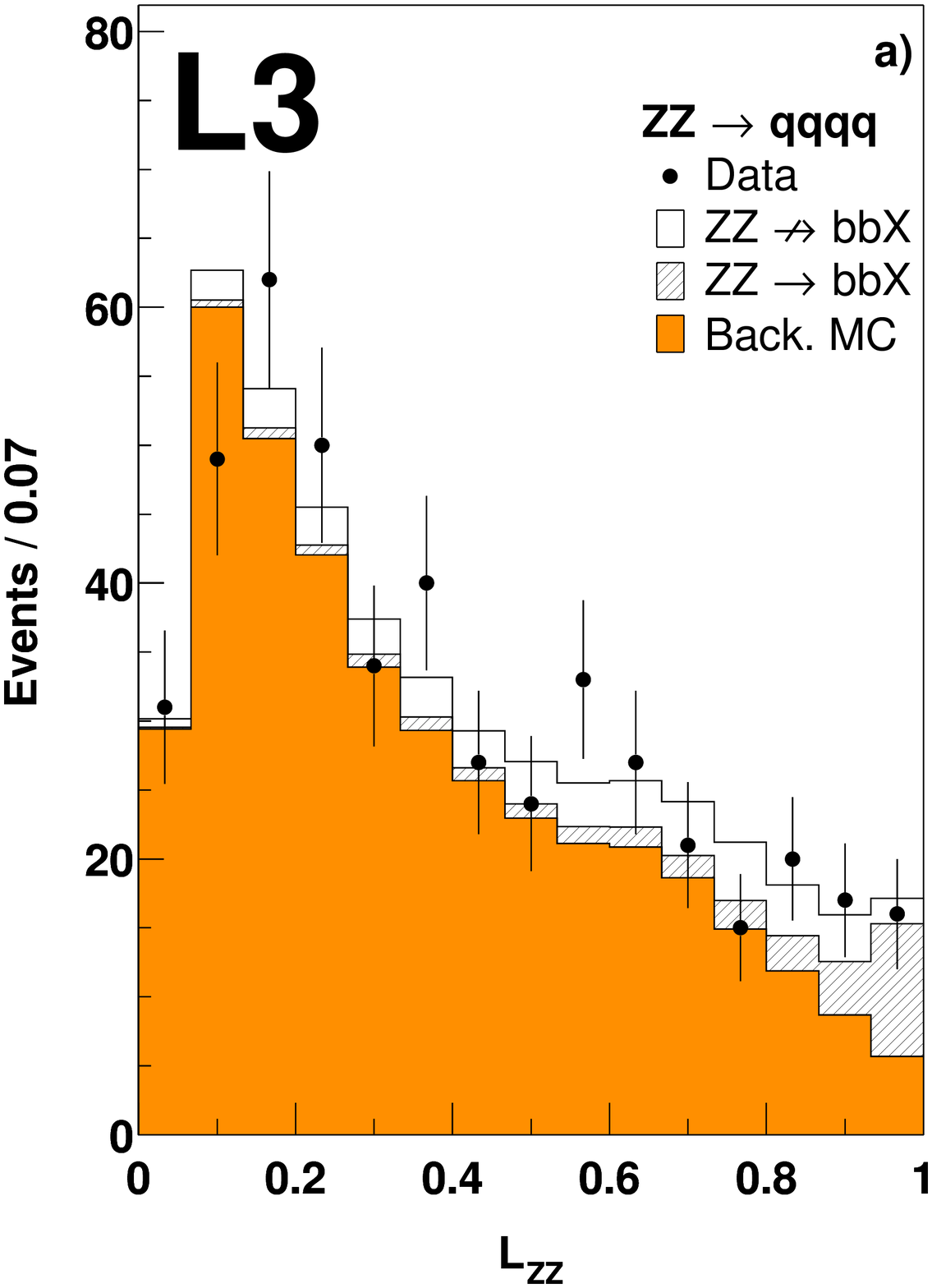} &
\includegraphics*[width=0.4\textwidth]{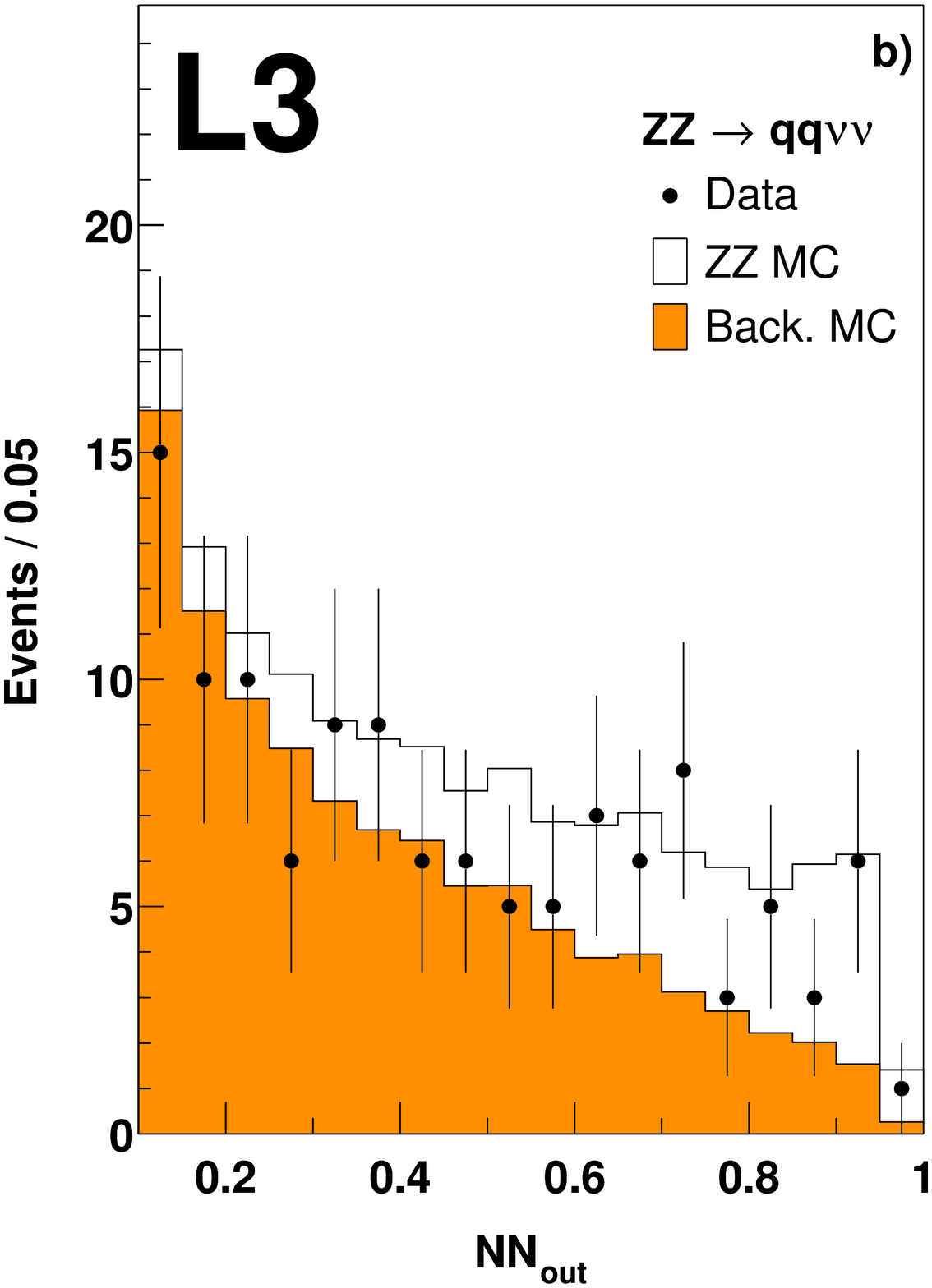} \\
\includegraphics*[width=0.4\textwidth]{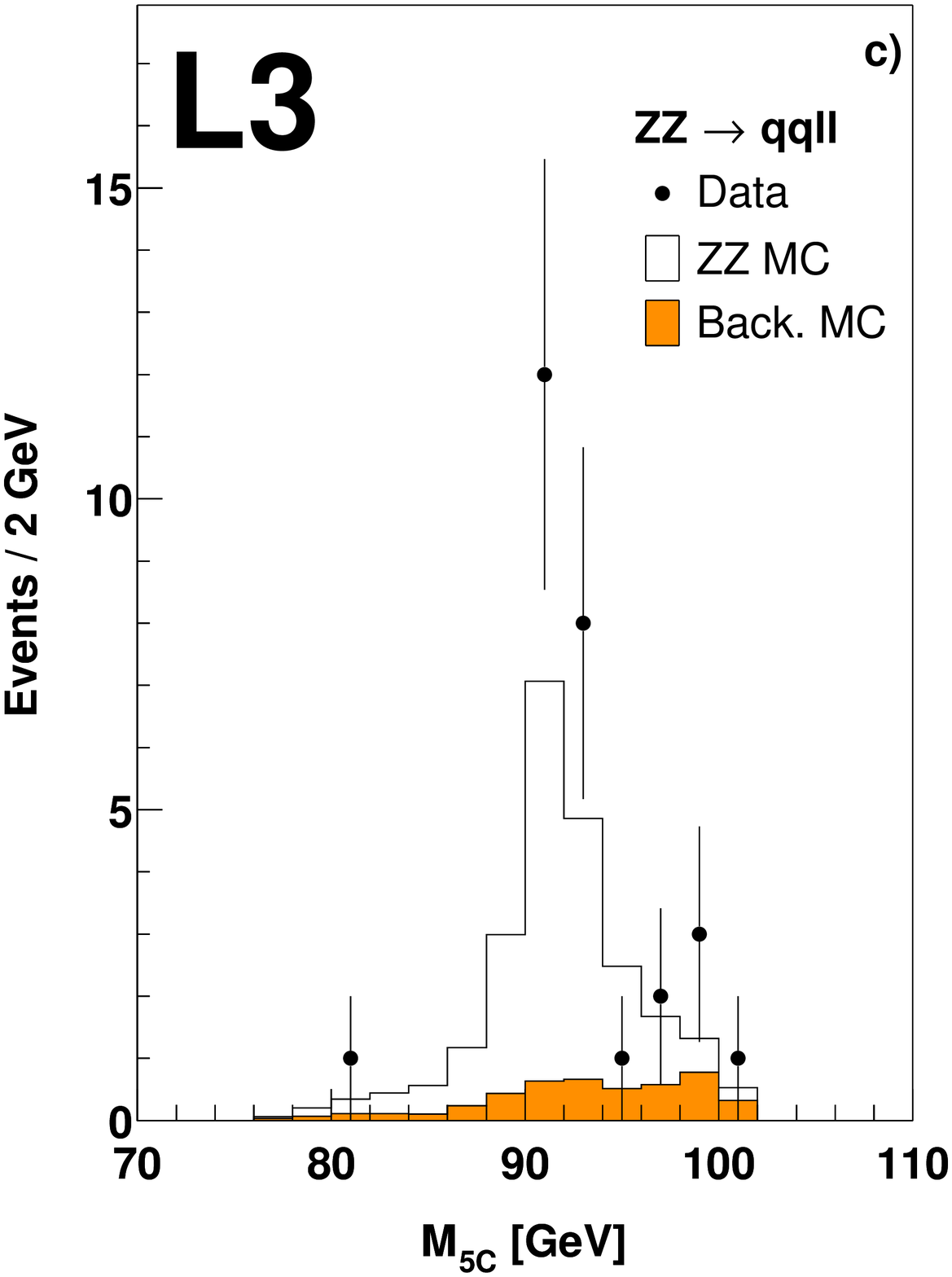} &
\includegraphics*[width=0.4\textwidth]{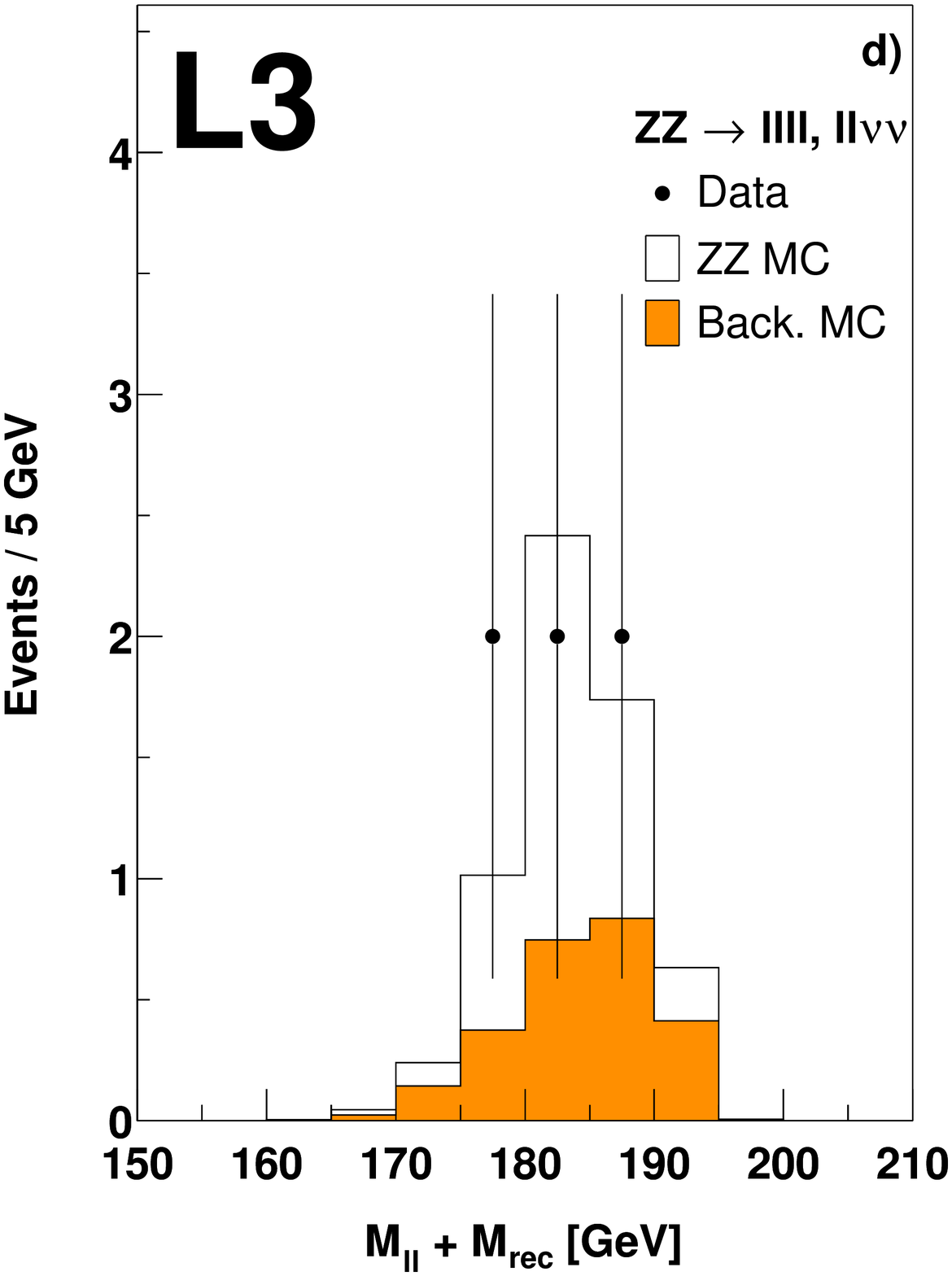} \\
\end{tabular}
\end{center}
\caption{Distribution of the final variables used for the
measurement of the cross-section for the a) \qqqq{} b) \qqnn{} and c)
\qqll{} final states. The sum of the \llnn{} and \llll{} final states
is given in d). The signal and background Monte Carlo
  distributions are normalised to the expected cross sections.}
\label{fig:finalvar}
\end{figure}

The distributions of $L_{\rm ZZ}$ for data and \MC{} are shown in
Figure~\ref{fig:finalvar}a. Table~\ref{tab:1} lists the yield of this
selection for $L_{\rm ZZ} > 0.2$, which corresponds to an efficiency
of 55.4\%.

%
%%%%%%%%%%%%%%%%%%%%%%%%%%%%%%%%%%%%%%%%%%%%%%%%%%%%%%%%%%%%%%%%%%%%%%
\subsection{The \boldmath{\qqnn{}} channel}
%%%%%%%%%%%%%%%%%%%%%%%%%%%%%%%%%%%%%%%%%%%%%%%%%%%%%%%%%%%%%%%%%%%%%%
%

The selection of the $\qqnn$ channel is identical to that performed at
$\sqrt{s}=192-202\GeV$~\cite{l3zz196}. High multiplicity events with
large missing energy and momentum and no high energy isolated
electrons, muons or photons are selected. They are forced into two
jets and a constrained fit which enforces the hypothesis that the
missing four-momentum is due to a Z boson is applied.

Finally, an artificial neural network  singles out Z
pair-production events from background.  Its input variables include event shape
variables that differentiate a two-jet from a three-jet topology, the
sum of visible and missing masses, the masses of the two jets, the
missing momentum and the energy in a 25$^\circ$ azimuthal sector
around the missing momentum vector.

Figure~\ref{fig:finalvar}b shows the output of the neural network,
$NN_{out}$, for data and \MC{}. The results of
this selection are summarised in Table~\ref{tab:1} for a benchmark cut
$NN_{out} > 0.5$, which corresponds to an efficiency of 46.2\%.

%
%%%%%%%%%%%%%%%%%%%%%%%%%%%%%%%%%%%%%%%%%%%%%%%%%%%%%%%%%%%%%%%%%%%%%%
\subsection{The \boldmath{\qqll{}} channel}
%%%%%%%%%%%%%%%%%%%%%%%%%%%%%%%%%%%%%%%%%%%%%%%%%%%%%%%%%%%%%%%%%%%%%%
%

The study of the \qqee, \qqmm\ and \qqtt\ final states~\cite{gagan-thesis} proceeds from a
sample of high multiplicity events, well balanced in the planes parallel
and transverse to the beam direction. Background from two-photon
interactions is rejected by requiring $|\cos \theta_T| <0.98$. Events
from the $\rm e^+e^- \rightarrow q\bar{q}(\gamma)$ process with hard
initial state radiation photons are reduced by requiring the effective
centre-of-mass energy~\cite{sprime}, $\sqrt{s'}$, to be greater than
$0.55 \sqrt{s}$. Remaining two-jet events are suppressed by a cut on
the event thrust.

To fully reconstruct the \qqll\ final state, after identifying an
electron, muon or tau pair in the event, the remaining clusters
are forced into two jets. A kinematic fit which enforces energy and
momentum conservation and equal mass, $M_{5C}$, of the hadronic
and leptonic systems is performed. To cope with different 
background contributions, different selection criteria are
applied for the three final states.

The \qqee\ selection requires low transverse energy imbalance and a
sum of the energies of the two electrons close to $\sqrt{s}/2$. The
variable $(E_1\,{+}\,E_2\,{-}\,E_3\,{-}\,E_4)/(E_1\,{+}\,E_2\,{+}\,E_3\,{+}\,E_4)$, is also
considered, where $E_i$ denotes the decreasingly ordered jet and
lepton energies. This variable has low values for the signal, where the
energy is uniformly distributed among the four particles, and large
values for the background from the $\rm e^+e^- \rightarrow
q\bar{q}(\gamma)$ process. An efficiency of 73.9\% is reached.

The \qqmm\ final state has a low background contamination, almost
entirely rejected by requiring a large energy for the lowest energetic
muon, expected to be soft for background events. This selection has an
efficiency of 60.4\%.

The \qqtt\ selection is affected by a larger background. It requires
the invariant mass of the hadronic system prior to the kinematic
fit to be compatible with $m_{\rm Z}$ and a large rest frame angle
between the taus. In addition, the sum of the di-jet and di-tau masses
after the kinematic fit has to be close to $2m_{\rm Z}$. This
selection accepts 28.0\% of the \qqtt\ final states, as well as 2.4\% and 4.8\% of the \qqee\ and \qqmm\ final states, respectively.

Table~\ref{tab:1} presents the combined yield of all selections, whose
overall efficiency is 55.4\%. Figure~\ref{fig:finalvar}c shows the $M_{5C}$
distributions for data and \MC{}.

%
%%%%%%%%%%%%%%%%%%%%%%%%%%%%%%%%%%%%%%%%%%%%%%%%%%%%%%%%%%%%%%%%%%%%%%
\subsection{The \boldmath{\llnn{}} channel}
%%%%%%%%%%%%%%%%%%%%%%%%%%%%%%%%%%%%%%%%%%%%%%%%%%%%%%%%%%%%%%%%%%%%%%
%

Only final states with
electrons and muons are considered with a selection identical to that
performed at lower $\sqrt{s}$~\cite{l3zz196}. A pair of acoplanar
leptons is selected in low multiplicity events with large missing
momentum pointing away from the beam axis. To improve efficiency,
electrons are not required to have an associated track. To reduce the
background, no MIP candidates are accepted in the muon selection. Both
the lepton visible mass, $M_{\ell\ell}$, and recoil mass, $M_{rec}$,
must be consistent with $m_{\rm Z}$. These criteria suppress
background from fermion pair-production. Residual background from
four-fermion processes is reduced by performing a fit which constrains
the leptons to originate from a Z boson and
requiring the recoil mass to be close to $m_{\rm Z}$.

For signal events, the sum $M_{\ell\ell}+M_{rec}$ should be close to
$2 m_{\rm Z}$.  The distributions of this variable for data and Monte
Carlo, combined with results from the \llll{} channel, are shown in
Figure~\ref{fig:finalvar}d. An efficiency of 25.3\% is achieved and
the results of the selection are reported in Table~\ref{tab:1}.

%
%%%%%%%%%%%%%%%%%%%%%%%%%%%%%%%%%%%%%%%%%%%%%%%%%%%%%%%%%%%%%%%%%%%%%%
\subsection{The \boldmath{\llll{}} channel}
%%%%%%%%%%%%%%%%%%%%%%%%%%%%%%%%%%%%%%%%%%%%%%%%%%%%%%%%%%%%%%%%%%%%%%
%

The selection for the \llll{} channel aims to retain a high efficiency
to compensate for the low branching ratio.  The same criteria of
Reference~\citen{l3zz196} are applied to select low multiplicity
events with four or more loosely identified leptons, with energy above
$3\GeV$. Events must contain at least one electron or muon pair.
Electrons without an associated track are accepted while MIPs are not
considered to form these pairs.

The lepton pair with mass closest to $m_{\rm Z}$ is selected and both
$M_{\ell\ell}$ and $M_{rec}$ are required to be compatible with
$m_{\rm Z}$. Background from fermion pair-production is reduced by
imposing upper bounds on the opening angle of the leptons of this
pair.

The sum $M_{\ell\ell}+M_{rec}$ is used as a final discriminating
variable. Its data and \MC{} distributions are presented in
Figure~\ref{fig:finalvar}d, together with those from the \llnn{}
channel.  The yield of the selection is summarized in
Table~\ref{tab:1} and corresponds to an efficiency of 33.6\%.

%
%%%%%%%%%%%%%%%%%%%%%%%%%%%%%%%%%%%%%%%%%%%%%%%%%%%%%%%%%%%%%%%%%%%%%%
\section{Results}
%%%%%%%%%%%%%%%%%%%%%%%%%%%%%%%%%%%%%%%%%%%%%%%%%%%%%%%%%%%%%%%%%%%%%%
%

%
%%%%%%%%%%%%%%%%%%%%%%%%%%%%%%%%%%%%%%%%%%%%%%%%%%%%%%%%%%%%%%%%%%%%%%
\subsection{Measurement of the ZZ cross section}
%%%%%%%%%%%%%%%%%%%%%%%%%%%%%%%%%%%%%%%%%%%%%%%%%%%%%%%%%%%%%%%%%%%%%%
%

The cross sections for each energy bin and each final state are
derived~\cite{l3zz196} with a fit to the final discriminating
variables and are presented in Table~\ref{tab:1} together with the
\SM{} predictions. In the presence of fluctuations for channels with low
statistics, an upper limit~\cite{l3zz196} on the cross section is
given. Fixing the relative contributions of all channels to the \SM{}
expectations, the Z pair-production cross section is extracted and
also presented in Table~\ref{tab:1}.  All the measured cross-sections
agree with their \SM{} predictions.

%
%%%%%%%%%%%%%%%%%%%%%%%%%%%%%%%%%%%%%%%%%%%%%%%%%%%%%%%%%%%%%%%%%%%%%%
\subsection{Study of systematic uncertainties}
%%%%%%%%%%%%%%%%%%%%%%%%%%%%%%%%%%%%%%%%%%%%%%%%%%%%%%%%%%%%%%%%%%%%%%
%

Several sources of systematic uncertainty are considered~\cite{l3zz196} and listed in
Table~\ref{tab:2}. Systematic effects correlated among channels arise from
uncertainties on the detector energy scales, on the signal modelling, as
derived from a comparison between EXCALIBUR and GRC4F and on the
prediction of the background level. This is studied by varying the expected 
cross sections for W pair-production, jet production, the $\rm
e\nu_\e q\bar q'$ and
four-fermion processes
by 0.5\%, 5\%, 10\% and 5\%, respectively. The $\qqqq$ channel is affected
by uncertainties on the charge multiplicity and the simulation of the
b-tag discriminant. Other sources of systematic uncertainty,
uncorrelated among the channels, are the signal and background Monte
Carlo statistics, detailed in Table~\ref{tab:3} and the accuracy of
the simulation of the selection variables and of those used for the
lepton identification.

\begin{figure}
\centerline{\includegraphics*[width=0.9\textwidth]{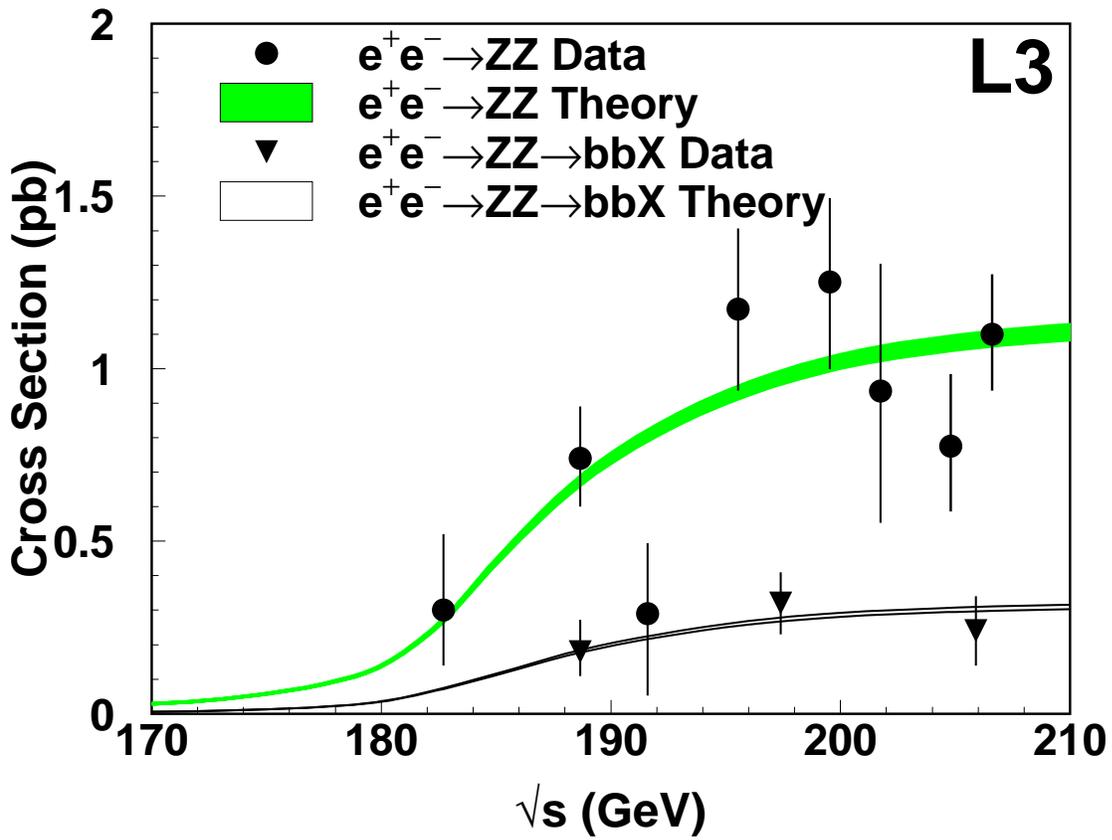}}
\caption{Measurements and predictions for the $\rm\eeto ZZ$ and $\rm
  \eeto ZZ\ra b\bar{b}X$
cross sections as a function of $\sqrt{s}$. An uncertainty of 2\% is assigned to the predictions.
}
\label{fig:xs}
\end{figure}

Including all systematic uncertainties, the
measured cross sections read:\footnote{In the NC02 framework, the cross sections are derived as:
\begin{eqnarray*}
  \nonumber
  \begin{array}{rcll}
    \sigma_{\ZZ}^{\mathrm{NC02}} (205 \mbox{\,GeV}) & = & 0.77 \pm 0.20 \mbox{\,(stat.)} \pm 0.05 \mbox{\,(syst.)} & \mbox{(SM :\,} 1.05 \pm 0.02 \mbox{\,pb)} \\
    \sigma_{\ZZ}^{\mathrm{NC02}} (207 \mbox{\,GeV}) & = & 1.09 \pm 0.17 \mbox{\,(stat.)} \pm 0.07 \mbox{\,(syst.)} & \mbox{(SM :\,} 1.07 \pm 0.02 \mbox{\,pb)},
  \end{array}
\end{eqnarray*}
where the Standard Model expectations, consistent among the 
EXCALIBUR, ZZTO~\cite{zzto} and YFSZZ~\cite{yfszz} programs, are
assigned an uncertainty of 2\%.} 
\begin{eqnarray*}
  \nonumber
  \begin{array}{rcll}
    \sigma_{\ZZ} (205 \mbox{\,GeV}) & = & 0.78 \pm 0.20 \mbox{\,(stat.)} \pm 0.05 \mbox{\,(syst.)} & \mbox{(SM :\,} 1.07 \pm 0.02 \mbox{\,pb)} \\
    \sigma_{\ZZ} (207 \mbox{\,GeV}) & = & 1.10 \pm 0.17 \mbox{\,(stat.)} \pm 0.07 \mbox{\,(syst.)} & \mbox{(SM :\,} 1.08 \pm 0.02 \mbox{\,pb)},
  \end{array}
\end{eqnarray*}
Figure~\ref{fig:xs} presents these values together with lower energy
measurements~\cite{l3zz183,l3zz189,l3zz196}.

%
%%%%%%%%%%%%%%%%%%%%%%%%%%%%%%%%%%%%%%%%%%%%%%%%%%%%%%%%%%%%%%%%%%%%%%
\subsection{Final states with b quarks}
%%%%%%%%%%%%%%%%%%%%%%%%%%%%%%%%%%%%%%%%%%%%%%%%%%%%%%%%%%%%%%%%%%%%%%
%

\begin{figure}
\centerline{\includegraphics*[angle=90,width=0.8\textwidth]{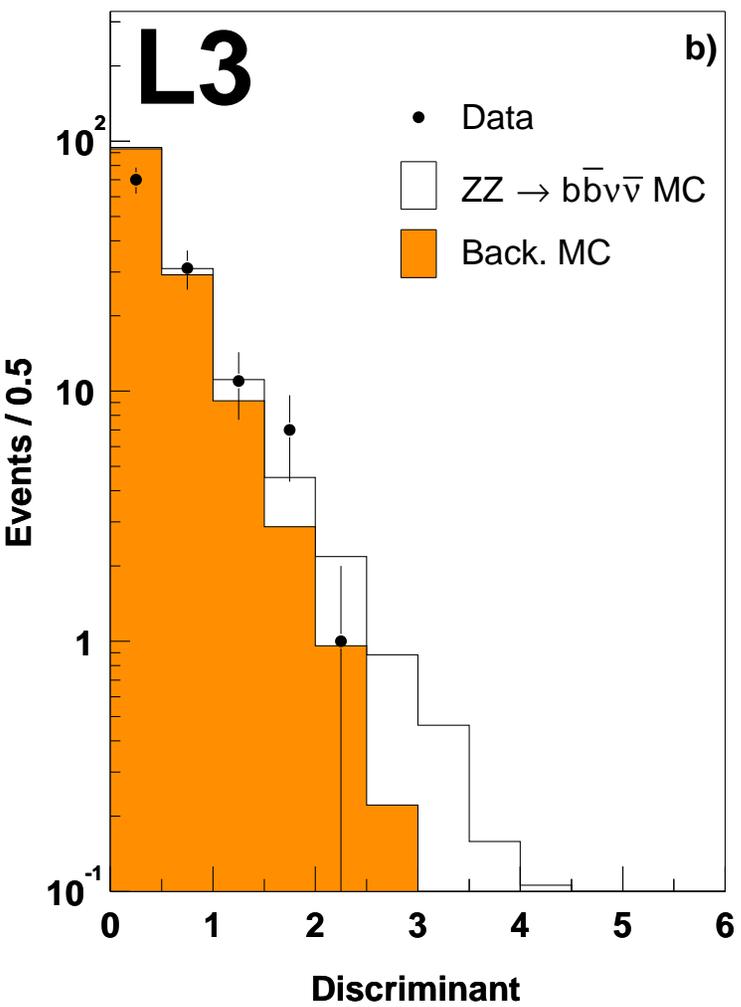}}
\centerline{\includegraphics*[angle=90,width=0.8\textwidth]{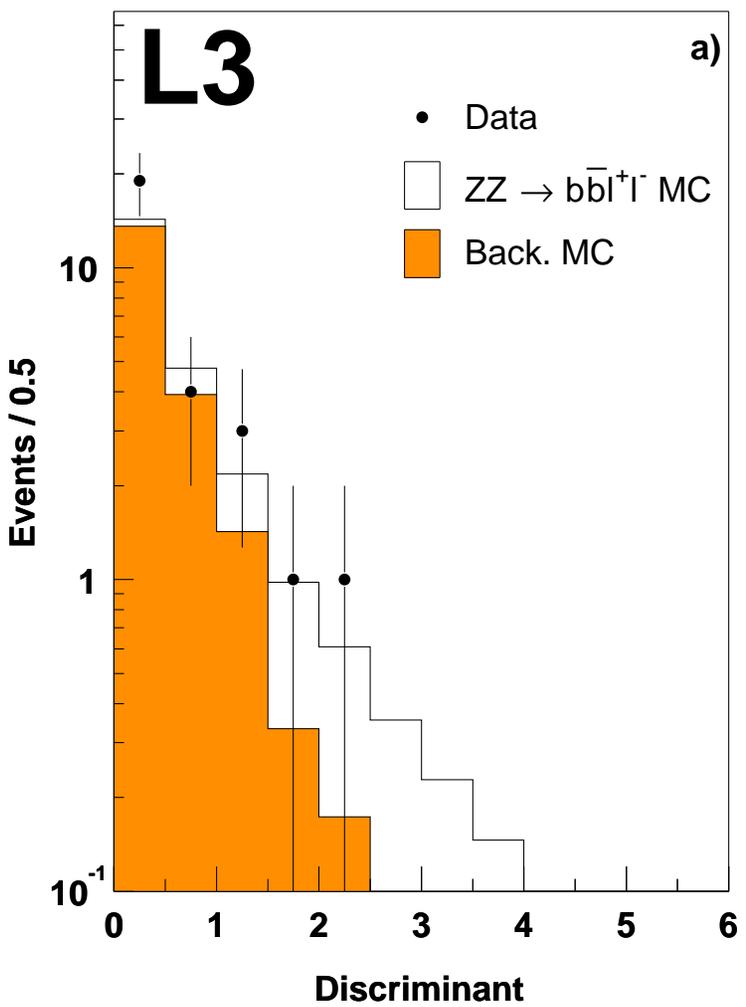}}
\caption{Discriminant variables in data and Monte Carlo for a) the
$\rm b\bar{b}\ell^+\ell^-$ and b) the $\rm b\bar{b}\nu\bar{\nu}$
selections. The signal and background Monte Carlo
  distributions are normalised to the expected cross sections. }
\label{fig:clbtag}
\end{figure}

The final discriminant of the \qqqq\ analysis, plotted in
Figure~\ref{fig:finalvar}a, shows a high sensitivity to final states
containing b quarks. In order to tag \bbnn\ and \bbll\ final states,
the b-tag information of the two jets are combined~\cite{l3zz196} to
form the discriminating variables shown in Figure~\ref{fig:clbtag}. For
the \bbnn\ final state, the value of the variable $NN_{out}$ is also
considered in the combination.
The cross sections for Z pair-production with
b quarks in the final states are determined from a fit to these
variables and listed in Table~\ref{tab:4}. Their combination gives a total
cross section:
\begin{displaymath}
\sigma_{\ZZtobbX} \mbox{\,(205 -- 207 GeV)}  = 0.24 \pm 0.09 \mbox{\,(stat)} \pm 0.03 \mbox{\,(sys)}.
\end{displaymath}
The systematic uncertainty
follows from the sources discussed above and is detailed in Table~\ref{tab:2}.

%
%%%%%%%%%%%%%%%%%%%%%%%%%%%%%%%%%%%%%%%%%%%%%%%%%%%%%%%%%%%%%%%%%%%%%%
\subsection{Combined results}
%%%%%%%%%%%%%%%%%%%%%%%%%%%%%%%%%%%%%%%%%%%%%%%%%%%%%%%%%%%%%%%%%%%%%%
%

The ratio between measured and expected cross sections, $R_{\rm ZZ}=\sigma^{fit}/\sigma^{th}$,
 is calculated including lower energy
 data~\cite{l3zz183,l3zz189,l3zz196}  as:
\begin{eqnarray*}
R_{\rm ZZ} \mbox{\,(183 -- 209 GeV)} & = & 0.93 \pm 0.08 \mbox{\,(stat)} \pm 0.06 \mbox{\,(sys)}.
\end{eqnarray*}
Systematic uncertainties include correlations among different data
samples.
The predictions are in agreement with the measurements with a
precision of 11\%.

\begin{figure}
\centerline{\includegraphics*[angle=90,width=0.8\textwidth]{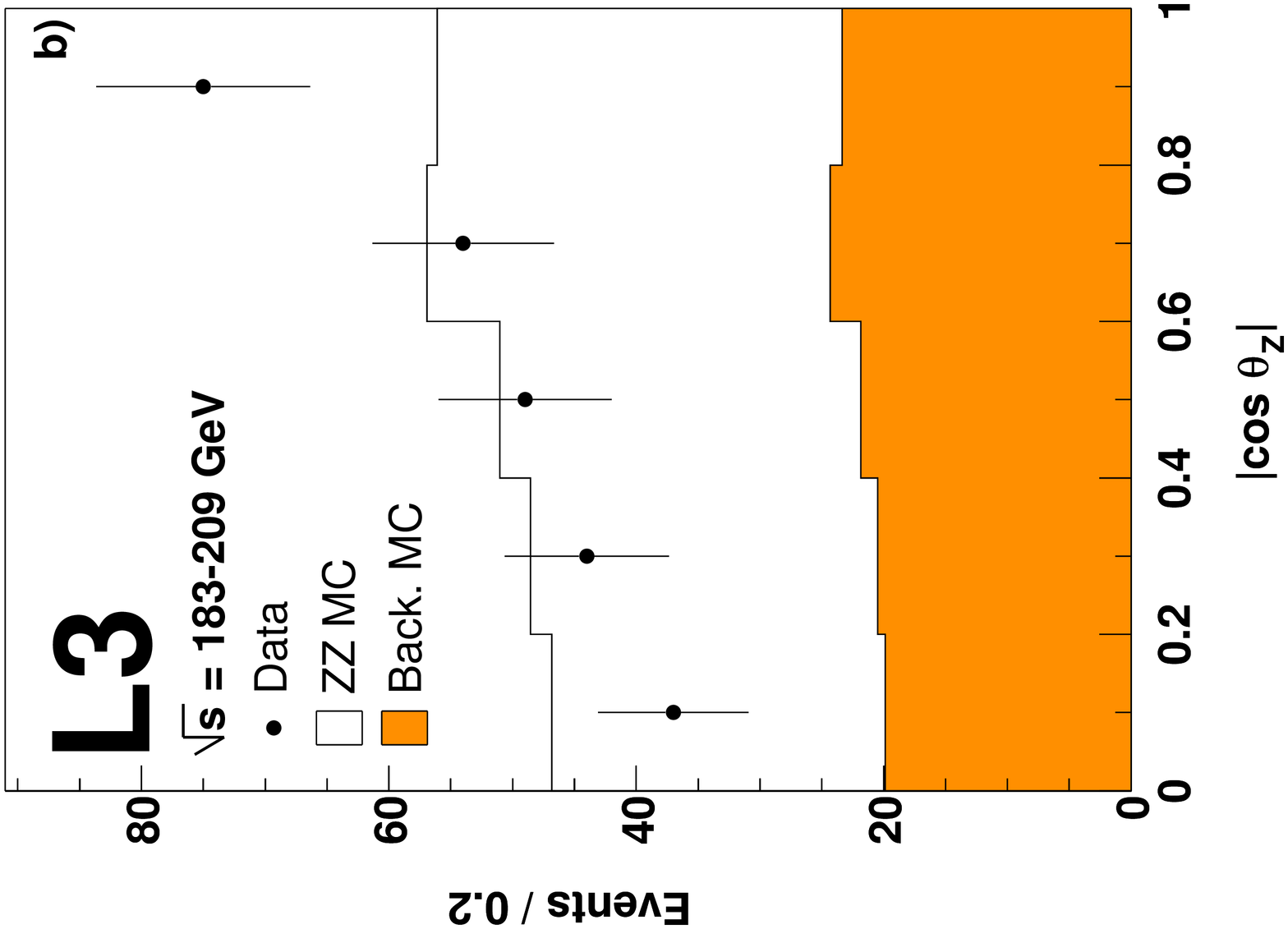}}
\centerline{\includegraphics*[angle=90,width=0.8\textwidth]{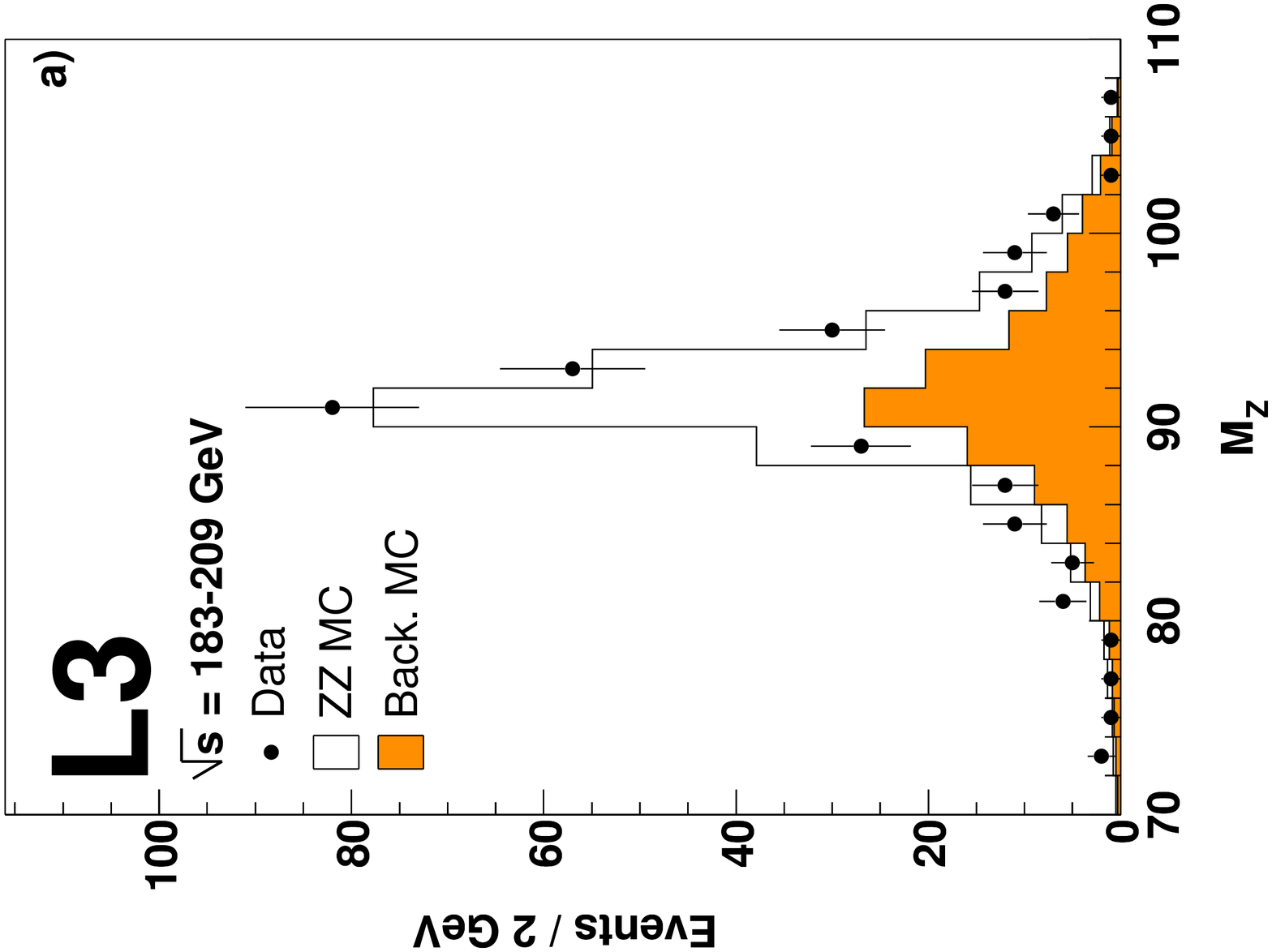}}
\caption{Distributions in data and \MC{} at all LEP centre-of-mass
energies above the Z pair production threshold of a) the reconstructed
mass $M$ and b) the absolute value of the cosine of the production
angle $\theta_{\rm Z}$. Cuts on
the  $\qqqq$ final discriminant and on the \qqnn{} neural network output
are applied.  The signal and background Monte Carlo
  distributions are normalised to the expected cross sections.}
\label{fig:massangle}
\end{figure}

Figure~\ref{fig:massangle}a shows the distribution of the
reconstructed mass $M_{\rm Z}$ of the Z boson, and Figure~\ref{fig:massangle}b the absolute value of the
cosine of the observed production angle $\theta_{\rm Z}$ for the full Z pair-production sample. The cuts
$L_{\rm ZZ} > 0.85$ and $NN_{out}>0.8$ are applied to data decsribed in
this Letter. Data at lower
energies~\cite{l3zz183,l3zz189,l3zz196} are also included.

%
%%%%%%%%%%%%%%%%%%%%%%%%%%%%%%%%%%%%%%%%%%%%%%%%%%%%%%%%%%%%%%%%%%%%%%
\section{Limits on physics beyond the Standard Model}
%%%%%%%%%%%%%%%%%%%%%%%%%%%%%%%%%%%%%%%%%%%%%%%%%%%%%%%%%%%%%%%%%%%%%%
%

%
%%%%%%%%%%%%%%%%%%%%%%%%%%%%%%%%%%%%%%%%%%%%%%%%%%%%%%%%%%%%%%%%%%%%%%
\subsection{Anomalous couplings}
%%%%%%%%%%%%%%%%%%%%%%%%%%%%%%%%%%%%%%%%%%%%%%%%%%%%%%%%%%%%%%%%%%%%%%
%

\begin{figure}
\centerline{\includegraphics*[width=0.65\textwidth]{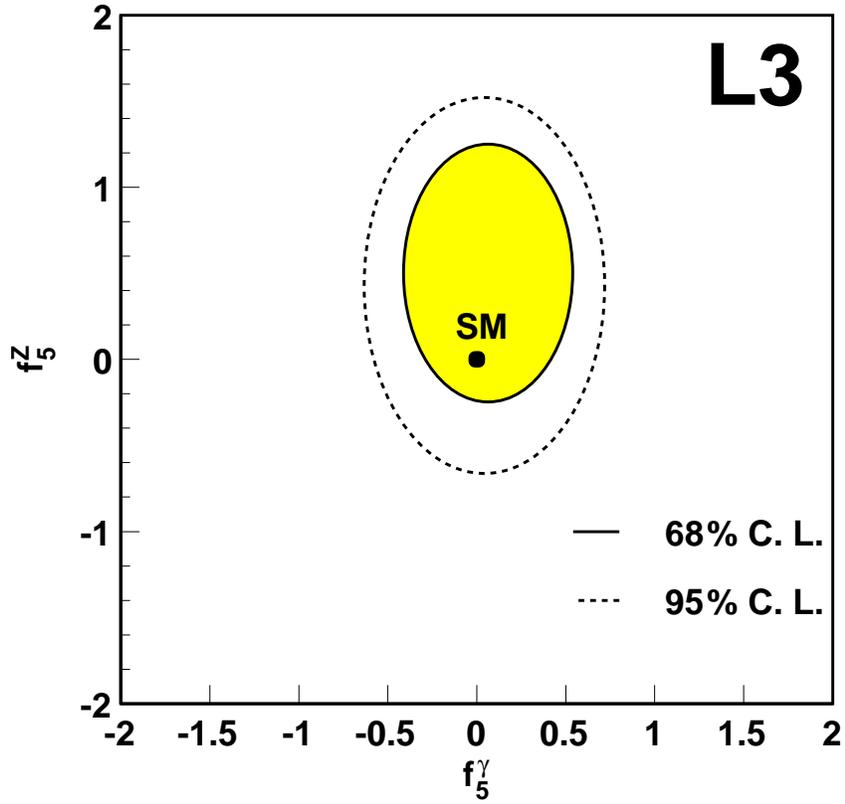}}
\vspace{5mm}
\centerline{\includegraphics*[width=0.65\textwidth]{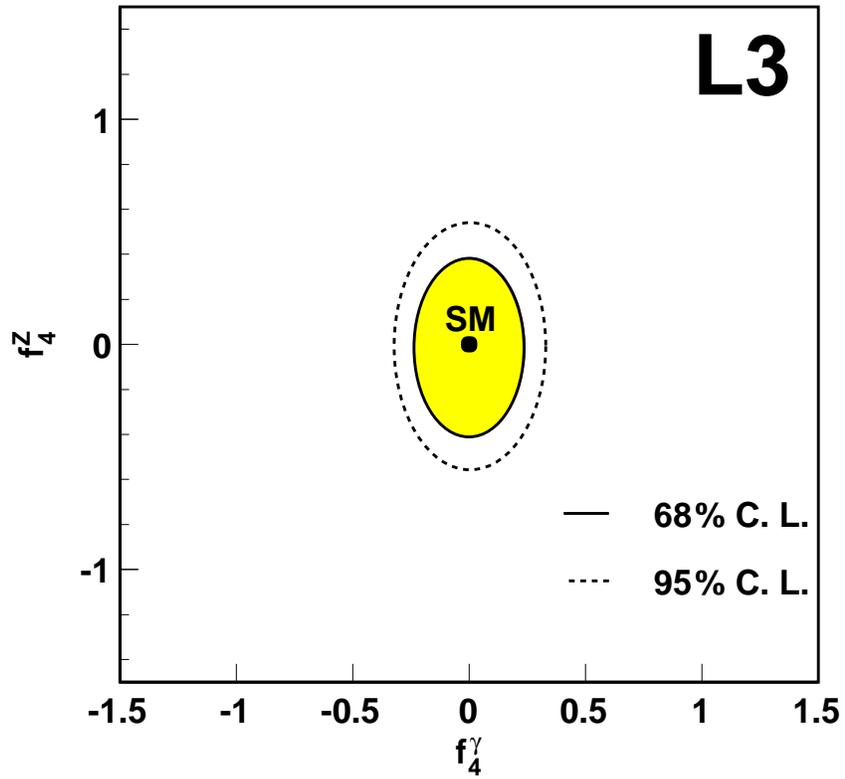}}
\caption{Results of a simultaneous fit to anomalous coupling
  parameters with the same CP eigenvalue. The Standard Model (SM)
  expectations are also indicated.}
\label{fig:anomalous}
\end{figure}

Assuming on-shell production of the two
Z bosons, anomalous ZZV couplings are parametrised~\cite{hagiwara} 
by the coefficients $f_i^{\mathrm V}$, with $i=4,5$ and $\rm
V=\gamma,Z$. The $f_4^{\rm V}$ coefficients correspond to CP violation
and the $f_5^{\rm V}$ ones to CP conservation. All $f_i^{\mathrm V}$
coefficients are zero in the Standard Model. Each event of the
signal Monte Carlo distributions presented in Figure~\ref{fig:finalvar} is
reweighted~\cite{l3zz189} to simulate anomalous values of the $f_i^{\mathrm
  V}$ coefficients. The full phase space of the Z boson pair, as
reconstructed from the jet and lepton four-momenta, is used. A fit to
these distribution is performed  
leaving one coupling free at a time and fixing the others to
zero, yielding the 95\% confidence level limits:
\begin{displaymath}
     -0.48 \leq f_4^{\Zo}       \leq 0.46 ;\hspace{2ex}
     -0.36 \leq f_5^{\Zo}       \leq 1.03 ;\hspace{2ex}
     -0.28 \leq f_4^{\gamma}    \leq 0.28 ;\hspace{2ex}
     -0.40 \leq f_5^{\gamma}    \leq 0.47,
\end{displaymath}
compatible with the Standard Model expectations. Lower energy data~\cite{l3zz183,l3zz189,l3zz196} are included in the
fit. Figure~\ref{fig:anomalous} presents results of  simultaneous fits to couplings with
the same CP eigenvalue.

%
%%%%%%%%%%%%%%%%%%%%%%%%%%%%%%%%%%%%%%%%%%%%%%%%%%%%%%%%%%%%%%%%%%%%%%
\subsection{Extra space dimensions}
%%%%%%%%%%%%%%%%%%%%%%%%%%%%%%%%%%%%%%%%%%%%%%%%%%%%%%%%%%%%%%%%%%%%%%
%

A recent theory~\cite{arkani}, dubbed ``Low Scale Gravity'', proposes
a solution to the hierarchy problem by postulating a scale $M_S$ for
the gravitational interactions which is of the order of the
electroweak scale.  Extra space dimensions are a consequence of this
theory.  In this scenario, spin-two gravitons
contribute to the Z pair-production~\cite{agashe}, interfering with
the Standard Model production mechanism.  The Z pair-production cross
sections presented in this Letter and those measured at lower
energies~\cite{l3zz183,l3zz189,l3zz196} are fit with a combination of
Low Scale Gravity and Standard Model contributions. A lower 95\%
confidence level limit on the scale $M_S$ of 0.7 TeV is obtained. It
holds for both constructive and destructive interference between the
Low Scale Gravity and the Standard Model contributions.

%
%%%%%%%%%%%%%%%%%%%%%%%%%%%%%%%%%%%%%%%%%%%%%%%%%%%%%%%%%%%%%%%%%%%%%%%%%%%%%%%
% Bibliography
%%%%%%%%%%%%%%%%%%%%%%%%%%%%%%%%%%%%%%%%%%%%%%%%%%%%%%%%%%%%%%%%%%%%%%%%%%%%%%
%
% Style file to use with mcite.
% Use l3style with just cite.
\bibliographystyle{l3stylem}

%
%%%%%%%%%%%%%%%%%%%%%%%%%%%%%%%%%%%%%%%%%%%%%%%%%%%%%%%%%%%%%%%%%%%%%%%%%%%%%%
% Author List
%%%%%%%%%%%%%%%%%%%%%%%%%%%%%%%%%%%%%%%%%%%%%%%%%%%%%%%%%%%%%%%%%%%%%%%%%%%%%%

\newpage
\typeout{   }     
\typeout{Using author list for paper 261 -  }
\typeout{$Modified: Jul 15 2001 by smele $}
\typeout{!!!!  This should only be used with document option a4p!!!!}
\typeout{   }
%
%
%
%  L A T E X  version!!
%
%
% Make sure that the Lep package has been used!
%\input{Lep.sty}%
%
%\ifx\LepCalled\undefined%
%\typeout{     }%
%\typeout{!!!!!!!!!!!!!!!!!!!!!!!!!!!!!!!!!!!!!!!!!!!!!!!!!!!!!!!!!!!}%
%\typeout{Yikes.  You haven't used the Lep package!}%
%\typeout{Please put \protect\usepackage\protect{Lep\protect} in your preamble,
%         followed by}%
%\typeout{\protect\Lep\protect{1\protect} or \protect\Lep\protect{2\protect}}%
%\typeout{     }%
%\typeout{For now you will get a Lep phase 2 authorlist (may not be right!).}%
%\typeout{!!!!!!!!!!!!!!!!!!!!!!!!!!!!!!!!!!!!!!!!!!!!!!!!!!!!!!!!!!!}%
%\typeout{     }%
%\Lep{2}\fi%

\newcount\tutecount  \tutecount=0
\def\tutenum#1{\global\advance\tutecount by 1 \xdef#1{\the\tutecount}}
\def\tute#1{$^{#1}$}
\tutenum\aachen            % 1
\tutenum\nikhef            % 2
\tutenum\mich              % 3
\tutenum\lapp              % 4
\tutenum\basel             % 5
\tutenum\lsu               % 6
\tutenum\beijing           % 7
\tutenum\bologna           % 8
\tutenum\tata              % 9 
\tutenum\ne                % 10
\tutenum\bucharest         % 11
\tutenum\budapest          % 12
\tutenum\mit               % 13
\tutenum\panjab            % 14 
\tutenum\debrecen          % 15
\tutenum\dublin            % 16
\tutenum\florence          % 17
\tutenum\cern              % 18
\tutenum\wl                % 19
\tutenum\geneva            % 20
\tutenum\hefei             % 21
\tutenum\lausanne          % 22
\tutenum\lyon              % 23
\tutenum\madrid            % 24
\tutenum\florida           % 25
\tutenum\milan             % 26
\tutenum\moscow            % 27
\tutenum\naples            % 29
\tutenum\cyprus            % 30
\tutenum\nymegen           % 31
\tutenum\caltech           % 32
\tutenum\perugia           % 33
\tutenum\peters            % 34
\tutenum\cmu               % 35
\tutenum\potenza           % 36
\tutenum\prince            % 37
\tutenum\riverside         % 38
\tutenum\rome              % 39
\tutenum\salerno           % 40
\tutenum\ucsd              % 41
\tutenum\sofia             % 42
\tutenum\korea             % 43
\tutenum\purdue            % 44
\tutenum\psinst            % 45
\tutenum\zeuthen           % 46
\tutenum\eth               % 47
\tutenum\hamburg           % 48
\tutenum\taiwan            % 49
\tutenum\tsinghua          % 50

{
\parskip=0pt
\noindent
{\bf The L3 Collaboration:}
\ifx\selectfont\undefined%  old style font selection
 \baselineskip=10.8pt
 \baselineskip\baselinestretch\baselineskip
 \normalbaselineskip\baselineskip
 \ixpt
\else%                      new style font selection
 \fontsize{9}{10.8pt}\selectfont
\fi
\medskip
\tolerance=10000
\hbadness=5000
\raggedright
\hsize=162truemm\hoffset=0mm
\def\r{\rlap,}
\noindent

P.Achard\r\tute\geneva\ 
O.Adriani\r\tute{\florence}\ 
M.Aguilar-Benitez\r\tute\madrid\ 
J.Alcaraz\r\tute{\madrid}\ 
G.Alemanni\r\tute\lausanne\
J.Allaby\r\tute\cern\
A.Aloisio\r\tute\naples\ 
M.G.Alviggi\r\tute\naples\
H.Anderhub\r\tute\eth\ 
V.P.Andreev\r\tute{\lsu,\peters}\
% ...
\newpage
%\rule{\textwidth}{0.4pt}
\begin{list}{A}{\itemsep=0pt plus 0pt minus 0pt\parsep=0pt plus 0pt minus 0pt
                \topsep=0pt plus 0pt minus 0pt}
\item[\aachen]
 III. Physikalisches Institut, RWTH, D-52056 Aachen, Germany$^{\S}$
\item[\nikhef] National Institute for High Energy Physics, NIKHEF, 
     and University of Amsterdam, NL-1009 DB Amsterdam, The Netherlands
\item[\mich] University of Michigan, Ann Arbor, MI 48109, USA
\item[\lapp] Laboratoire d'Annecy-le-Vieux de Physique des Particules, 
     LAPP,IN2P3-CNRS, BP 110, F-74941 Annecy-le-Vieux CEDEX, France
\item[\basel] Institute of Physics, University of Basel, CH-4056 Basel,
     Switzerland
\item[\lsu] Louisiana State University, Baton Rouge, LA 70803, USA
\item[\beijing] Institute of High Energy Physics, IHEP, 
  100039 Beijing, China$^{\triangle}$ 
\item[\bologna] University of Bologna and INFN-Sezione di Bologna, 
     I-40126 Bologna, Italy
\item[\tata] Tata Institute of Fundamental Research, Mumbai (Bombay) 400 005, India
\item[\ne] Northeastern University, Boston, MA 02115, USA
\item[\bucharest] Institute of Atomic Physics and University of Bucharest,
     R-76900 Bucharest, Romania
\item[\budapest] Central Research Institute for Physics of the 
     Hungarian Academy of Sciences, H-1525 Budapest 114, Hungary$^{\ddag}$
\item[\mit] Massachusetts Institute of Technology, Cambridge, MA 02139, USA
\item[\panjab] Panjab University, Chandigarh 160 014, India.
\item[\debrecen] KLTE-ATOMKI, H-4010 Debrecen, Hungary$^\P$
\item[\dublin] Department of Experimental Physics,
  University College Dublin, Belfield, Dublin 4, Ireland
\item[\florence] INFN Sezione di Firenze and University of Florence, 
     I-50125 Florence, Italy
\item[\cern] European Laboratory for Particle Physics, CERN, 
     CH-1211 Geneva 23, Switzerland
\item[\wl] World Laboratory, FBLJA  Project, CH-1211 Geneva 23, Switzerland
\item[\geneva] University of Geneva, CH-1211 Geneva 4, Switzerland
\item[\hefei] Chinese University of Science and Technology, USTC,
      Hefei, Anhui 230 029, China$^{\triangle}$
\item[\lausanne] University of Lausanne, CH-1015 Lausanne, Switzerland
\item[\lyon] Institut de Physique Nucl\'eaire de Lyon, 
     IN2P3-CNRS,Universit\'e Claude Bernard, 
     F-69622 Villeurbanne, France
\item[\madrid] Centro de Investigaciones Energ{\'e}ticas, 
     Medioambientales y Tecnol\'ogicas, CIEMAT, E-28040 Madrid,
     Spain${\flat}$ 
\item[\florida] Florida Institute of Technology, Melbourne, FL 32901, USA
\item[\milan] INFN-Sezione di Milano, I-20133 Milan, Italy
\item[\moscow] Institute of Theoretical and Experimental Physics, ITEP, 
     Moscow, Russia
\item[\naples] INFN-Sezione di Napoli and University of Naples, 
     I-80125 Naples, Italy
\item[\cyprus] Department of Physics, University of Cyprus,
     Nicosia, Cyprus
\item[\nymegen] University of Nijmegen and NIKHEF, 
     NL-6525 ED Nijmegen, The Netherlands
\item[\caltech] California Institute of Technology, Pasadena, CA 91125, USA
\item[\perugia] INFN-Sezione di Perugia and Universit\`a Degli 
     Studi di Perugia, I-06100 Perugia, Italy   
\item[\peters] Nuclear Physics Institute, St. Petersburg, Russia
\item[\cmu] Carnegie Mellon University, Pittsburgh, PA 15213, USA
\item[\potenza] INFN-Sezione di Napoli and University of Potenza, 
     I-85100 Potenza, Italy
\item[\prince] Princeton University, Princeton, NJ 08544, USA
\item[\riverside] University of Californa, Riverside, CA 92521, USA
\item[\rome] INFN-Sezione di Roma and University of Rome, ``La Sapienza",
     I-00185 Rome, Italy
\item[\salerno] University and INFN, Salerno, I-84100 Salerno, Italy
\item[\ucsd] University of California, San Diego, CA 92093, USA
\item[\sofia] Bulgarian Academy of Sciences, Central Lab.~of 
     Mechatronics and Instrumentation, BU-1113 Sofia, Bulgaria
\item[\korea]  The Center for High Energy Physics, 
     Kyungpook National University, 702-701 Taegu, Republic of Korea
\item[\purdue] Purdue University, West Lafayette, IN 47907, USA
\item[\psinst] Paul Scherrer Institut, PSI, CH-5232 Villigen, Switzerland
\item[\zeuthen] DESY, D-15738 Zeuthen, Germany
\item[\eth] Eidgen\"ossische Technische Hochschule, ETH Z\"urich,
     CH-8093 Z\"urich, Switzerland
\item[\hamburg] University of Hamburg, D-22761 Hamburg, Germany
\item[\taiwan] National Central University, Chung-Li, Taiwan, China
\item[\tsinghua] Department of Physics, National Tsing Hua University,
      Taiwan, China
\item[\S]  Supported by the German Bundesministerium 
        f\"ur Bildung, Wissenschaft, Forschung und Technologie
\item[\ddag] Supported by the Hungarian OTKA fund under contract
numbers T019181, F023259 and T037350.
\item[\P] Also supported by the Hungarian OTKA fund under contract
  number T026178.
\item[$\flat$] Supported also by the Comisi\'on Interministerial de Ciencia y 
        Tecnolog{\'\i}a.
\item[$\sharp$] Also supported by CONICET and Universidad Nacional de La Plata,
        CC 67, 1900 La Plata, Argentina.
\item[$\triangle$] Supported by the National Natural Science
  Foundation of China.
\end{list}
}
\vfill

%%% Local Variables: 
%%% mode: latex
%%% TeX-master: t
%%% End:

\newpage

%
%%%%%%%%%%%%%%%%%%%%%%%%%%%%%%%%%%%%%%%%%%%%%%%%%%%%%%%%%%%%%%%%%%%%%%%%%%%%%%
% Tables
%%%%%%%%%%%%%%%%%%%%%%%%%%%%%%%%%%%%%%%%%%%%%%%%%%%%%%%%%%%%%%%%%%%%%%%%%%%%%%
%

\begin{table}
  \begin{center}
    \begin{tabular}{|c|c|ccc|cc|}
      \hline
       \rule{0pt}{12pt}Channel & $\sqrt{s}$ (GeV) & $N_D$ & $N_S$ & $N_B$ & $\sigma^{fit}$ (pb) & $\sigma^{th}$ (pb) \\
 
      \hline
      \rule{0pt}{12pt}\qqqq     
      & 205 &    166   &     24.9 $\pm$ 0.0 & 140.5 $\pm$ 0.4       & $\phantom{<}0.38^{+0.20}_{-0.17}$ & 0.51\\ 
      \rule{0pt}{12pt}& 207 &    300   &     46.6 $\pm$ 0.0 & 255.1 $\pm$ 0.6       & $\phantom{<}0.55^{+0.15}_{-0.14}$ & 0.52\\ 
      \hline
      \rule{0pt}{12pt}\qqnn
      & 205 & \pho13   &     10.8 $\pm$ 0.1 & \pho11.0 $\pm$ 0.1    & $\hspace{-4pt}<0.24\phantom{^{+0.00}_{-0.00}}$ & 0.30\\
      \rule{0pt}{12pt}& 207 & \pho36   &     19.3 $\pm$ 0.1 & \pho18.7 $\pm$ 0.1    & $\phantom{<}0.25^{+0.08}_{-0.08}$ & 0.30\\
      \hline
      \rule{0pt}{12pt}\qqll     
      & 205 & \pho10   &  \pho6.8 $\pm$ 0.1 & \pho\pho1.6 $\pm$ 0.1 & $\phantom{<}0.19^{+0.08}_{-0.06}$ & 0.16\\  
      \rule{0pt}{12pt}& 207 & \pho18   &     12.3 $\pm$ 0.0 & \pho\pho3.0 $\pm$ 0.1 & $\phantom{<}0.19^{+0.06}_{-0.05}$ & 0.16\\  
      \hline
      \rule{0pt}{12pt}\llnn     
      & 205 &\pho\pho2 &  \pho0.9 $\pm$ 0.0 & \pho\pho0.6 $\pm$ 0.0 & $\phantom{<}0.08^{+0.09}_{-0.06}$ & 0.04\\
      \rule{0pt}{12pt}& 207 &\pho\pho3 &  \pho1.6 $\pm$ 0.1 & \pho\pho1.2 $\pm$ 0.0 & $\phantom{<}0.05^{+0.06}_{-0.04}$ & 0.04\\    
      \hline
      \rule{0pt}{12pt}\llll     
      & 205 &\pho\pho0 &  \pho0.4 $\pm$ 0.0 & \pho\pho0.2 $\pm$ 0.0 & $\hspace{-4pt}<0.11\phantom{^{+0.00}_{-0.00}}$ & 0.02\\
      \rule{0pt}{12pt}& 207 &\pho\pho1 &  \pho0.7 $\pm$ 0.0 & \pho\pho0.4 $\pm$ 0.0 & $\phantom{<}0.08^{+0.09}_{-0.06}$ & 0.02\\
      \hline
      \rule{0pt}{12pt}{$\rm e^+e^- \rightarrow ZZ$}     
      & 205 &      191 &     43.8 $\pm$ 0.1 & 154.0 $\pm$ 0.4       & $\phantom{<}0.78\pm0.20$          & 1.07\\
      \rule{0pt}{12pt}& 207 &      358 &     80.4 $\pm$ 0.1 & 278.4 $\pm$ 0.6       & $\phantom{<}1.10\pm0.17$          & 1.08\\
      \hline
    \end{tabular}
  \end{center}

  \caption{Number of observed data events, $N_D$ and signal, $N_S$, and background,
  $N_B$, expected Monte Carlo events in the two energy bins. The benchmark
  criteria $L_{\rm ZZ}>0.2$ and $NN_{Out}>0.5$ are applied for the
  $\qqqq$ and $\qqnn$ final states, respectively.  Uncertainties are
  due to \MC{} statistics. Measured, $\sigma^{fit}$, and expected,
  $\sigma^{th}$, cross sections are also given. Limits on
  $\sigma^{fit}$ are at the 95\% confidence level.}
  \label{tab:1}
\end{table}

\begin{table}
  \begin{center}
    \begin{tabular}{|c|c|c|}
        \hline
                                   & $\delta\sigma_{\rm ZZ}$ (\%) & $\delta\sigma_{\ZZtobbX}$ (\%) \\ 
        \hline
        Correlated sources         &     &     \\
        \hline
        Energy scale               & 3.1 & $\phantom{0}$2.4 \\
        Theory predictions         & 2.0 & $\phantom{0}$2.0 \\
        WW cross section           & 0.4 & $\phantom{0}$0.5 \\
        Four-jet rate              & 1.4 & $\phantom{0}$2.7 \\
        ${\rm W e \nu}$   cross section          & 1.1 & $\phantom{0}$0.8 \\ 
        Four-fermion cross section & 0.5 & $\phantom{0}$0.5 \\ 
        \hline
        Uncorrelated sources       & & \\
        \hline                     
        Charge multiplicity        & 1.3 & $\phantom{0}$2.3 \\
        B-tag                      & 2.5 & 11.3 \\
        Monte Carlo statistics     & 1.9 & $\phantom{0}$3.1 \\
        Simulation       & 3.5 & $\phantom{0}$3.5 \\
        \hline                        
        Total                      & 6.4 & 13.2 \\
        \hline
    \end{tabular}
  \end{center}

  \caption{Systematic uncertainties on $\sigma_{\rm ZZ}$ and
      $\sigma_{\ZZtobbX}$. }

  \label{tab:2}
\end{table}

\begin{table}
  \begin{center}
    \begin{tabular}{|c|c|c|c|c|c|}
        \hline
        & \qqll & \qqnn & \qqqq & \llnn & \llll \\
        \hline
        Signal MC statistics ($\sigma_{\rm ZZ}$)          & 0.5\,\% & 0.4\,\% & $<$0.1\,\% & 3.0\,\% & 1.0\,\%
        \\
        Background MC statistics ($\sigma_{\rm ZZ}$)      & 3.4\,\% & 0.5\,\% & $\phantom{<}$0.2\,\%    & 0.1\,\% & 0.2\,\%
        \\
        \hline
        Signal MC statistics ($\sigma_{\ZZtobbX}$)     & 3.9\,\% & 1.8\,\% & $\phantom{<}$2.1\,\%    & --       & --
        \\
        Background MC statistics ($\sigma_{\ZZtobbX}$) & 3.4\,\% & 1.4\,\% & $\phantom{<}$2.2\,\%    & --       & --
        \\
        \hline
        Simulation                          & 4.8\,\% & 3.2\,\% & $\phantom{<}$1.6\,\%    & 0.8\,\% & 1.8\,\%\\
        \hline
    \end{tabular}
  \end{center}

  \caption{Sources of uncorrelated systematic uncertainties on
      $\sigma_{\rm ZZ}$ and $\sigma_{\ZZtobbX}$.}

  \label{tab:3}

\end{table}
\begin{table}
  \begin{center}
    \begin{tabular}{|c|c|c|c|}
        \hline\rule{0pt}{14pt} 
        & \bbll & \bbnn & \qqbb \\
        \hline
        Measured cross section (pb) & 0.032 $\pm$ 0.027 & $\hspace{-4pt}<0.108$ & 0.185 $\pm$ 0.074 \\
        Expected cross section (pb) & 0.035             & $\phantom{<}0.065$    & 0.201 \\
        \hline
    \end{tabular}
  \end{center}
  \caption{Cross sections for final states with b quarks. The limit is at 95\% confidence level.}
  \label{tab:4}
\end{table}

%
%%%%%%%%%%%%%%%%%%%%%%%%%%%%%%%%%%%%%%%%%%%%%%%%%%%%%%%%%%%%%%%%%%%%%%%%%%%%%%
% Figures
%%%%%%%%%%%%%%%%%%%%%%%%%%%%%%%%%%%%%%%%%%%%%%%%%%%%%%%%%%%%%%%%%%%%%%%%%%%%%%
%

\end{document}